\documentclass{aa501}
 
\usepackage[latin1]{inputenc}
\usepackage{graphicx}
\usepackage{natbib}
\begin{document}
\title{Warm-hot intergalactic baryons revealed}
\author{L. Zappacosta\inst{1}, F. Mannucci\inst{2},
R. Maiolino\inst{3}, R. Gilli\inst{3},\\ A. Ferrara\inst{3},
A. Finoguenov\inst{4}, N. M. Nagar\inst{3}, D. J. Axon\inst{5}}
\institute{Dipartimento di Astronomia e Scienza dello Spazio, Largo
E. Fermi 2, I-50125 Firenze, Italy
\and Centro per l'Astronomia Infrarossa e lo studio del mezzo
interstellare (CAISMI) - CNR, Largo E.Fermi 5, I-50125 Firenze 
Italy 
\and Osservatorio Astrofisico di Arcetri, Largo E. Fermi 5, I-50125
Firenze, Italy 
\and
Max-Planck-Institut f\"ur extraterrestrische Physik,
             Giessenbachstra\ss e, 85748 Garching, Germany
\and
Department of Physical Sciences, University of Hertfordshire, College
Lane, Hatfield, Hertfordshire AL10 9AB, England, UK
}

\authorrunning{Zappacosta et al.}
\titlerunning{Warm-Hot Intergalactic Baryons}
\date{Received ..., Accepted ...}
\normalsize
\abstract{Several popular cosmological models predict that most of the baryonic mass 
in the local universe is located in filamentary and sheet-like structures 
associated with galaxy overdensities. 
This gas is expected to be gravitationally heated to $\sim 10^6 ~K$
and therefore emitting in the soft X-rays. 
We have detected diffuse soft X-ray structures in a
high Galactic latitude ROSAT field after point source subtraction and 
correction for Galactic absorption.
 These diffuse structures have an X-ray
energy distribution that is much softer than expected from clusters, groups or 
unresolved emission from AGNs, but are consistent with that 
expected from a diffuse warm intergalactic medium.
To discriminate between a Galactic or extragalactic nature of the
diffuse gas we have correlated the soft X--map with multiband optical 
images in this field. We have found a significant 
overdensity of galaxies in correspondence with the strongest diffuse 
X-ray structure. The
photometric redshift distribution of the galaxies over the X-ray peak has
an excess over field galaxies at $z\sim0.45$. This result
strongly suggests that the diffuse X-ray flux is due to extragalactic 
emission by
warm gas associated with an overdense galaxy region at $z\sim0.45$.
}
\maketitle

\section{Introduction}
The mismatch between the density of baryons observed in the local Universe
and the baryon density observed and predicted at high redshift is currently
one of the most puzzling issues in cosmology. While the
density of baryons, $\Omega _b$, observed in stars and gas ($\rm{HI}+
\rm{HII}$) in the local
Universe does not exceed 0.01 \citep{fukugita}, observations of the 
Ly$\alpha$ forest at $z=2$ \citep{rauch} and Big Bang nucleosynthesis 
constraints \citep{burles} both 
give $\Omega _b \simeq 0.04$ or larger \citep{pettini}.
One possibility is that at $z<1$ the baryonic gas falls onto the cosmic 
web pattern and is heated by shock mechanisms forming filamentary and 
sheet-like structures 
\citep{cen,dave}.
Such diffuse gas, called Warm--Hot Intergalactic Medium (WHIM),
should be detectable in the soft X-rays as a consequence of having a 
temperature in the range $\rm 10^5 ~K < \rm{T} < 10^7 ~K$.
A tight connection with the filamentary distribution of galaxies is expected
from N-body simulations (see \citet{bond} for a
theoretical discussion). 
\\
Various models of diffuse X-ray emission have been
proposed by several authors during the past year \citep{phillips,kuntz,bryan};
nonetheless, observational
evidence for diffuse/filamentary X-ray emission is still sparse.
\citet{wang} found an excess of emission
in some ROSAT fields which they ascribed to a diffuse component of the
X-ray background due to WHIM.
\citet{soltan} detected a correlation signature between
the soft X-ray background and galaxies.
Filamentary soft X-ray structures were identified
 by
\citet{warwick} in 
various overlapping ROSAT pointings near the Lockman Hole. 
Structures laid among clusters were found by \citet{kull} in the core
of the Shapley Supercluster  
and \citet{tittley} along the line of sight connecting two Abell clusters. 
\citet{scharf} presented tentative evidence of a soft X-ray
filamentary structure which 
seems to correlate with the density of galaxies measured in the I-band.
Recently, \citet{bagchi} pointed out the discovery of a presumably
filamentary structure both in radio and soft X-ray traced by a galaxy arc.

We have re-analyzed several ROSAT pointings toward a region close
to the Lockman Hole
with exposure times $> 10 \rm{~ksec}$ and, after
removal of the point sources and correction for Galactic absorption,
we have detected diffuse X-ray emission. We have started a program of optical
multiband wide field imaging of the regions showing diffuse X-ray emission
with the goal of detecting associated galaxy overdensities, which
would support the extragalactic nature of the large scale soft X-ray 
structures.

In this paper we report preliminary results on one of the ROSAT fields
 for which we have obtained HI radio data and which has been partially
 mapped in five optical bands. We present the discovery of a galaxy
 overdensity at a photometric redshift $z\sim$0.45 spatially
 coincident with the most prominent diffuse X-ray structure in the
 ROSAT field.\\ The plan of the paper is as follows: in
 Sect.\ref{xmaps} we discuss the analysis of the X-ray ROSAT maps, in
 Sect.\ref{HI} the reduction of the radio data and the HI absorption
 correction for X-ray maps are presented, in Sect.\ref{galaxies} we
 discuss the optical images analysis. Sect.\ref{photom z} contains the
 photometric redshift estimate and, finally, in Sect.\ref{summary}
 a summary of the results is presented.

\begin{figure*}[!]
\begin{center}      
\vskip2truecm
\caption{Panel {\em (a)} Soft (R2, $0.14-0.284 \rm{~keV}$) map smoothed 
to 15 arcsec showing the point sources; {\em (b)} Hard 
(R6+7, $0.73-2.04 \rm{~keV}$) map with the 
 same smoothing.
The inner circle (1~deg in diameter) encloses
the region where the point source
removal is reliable. The two crosses indicate the location of
the structures discussed in Fig.\ref{maps}.}
\label{maps_p}
\end{center}
\begin{picture}(100,-100)
\put(70,245){{\em (a)}}
\put(280,245){{\em (b)}}
\end{picture}
\end{figure*}
\begin{figure*}[t!]
\begin{center}
\includegraphics[angle=0, width=0.4\textwidth]{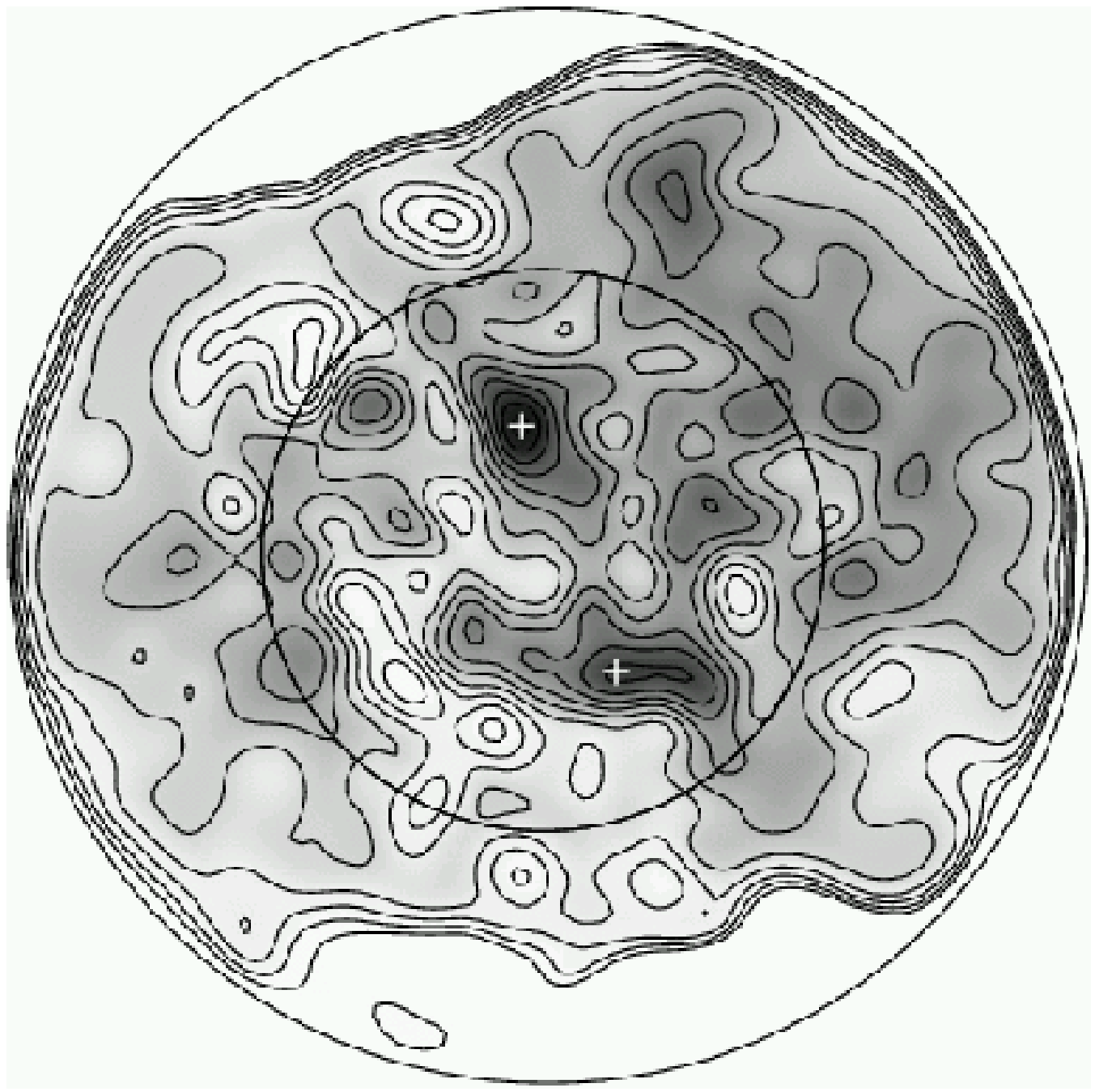}
\includegraphics[angle=0, width=0.4\textwidth]{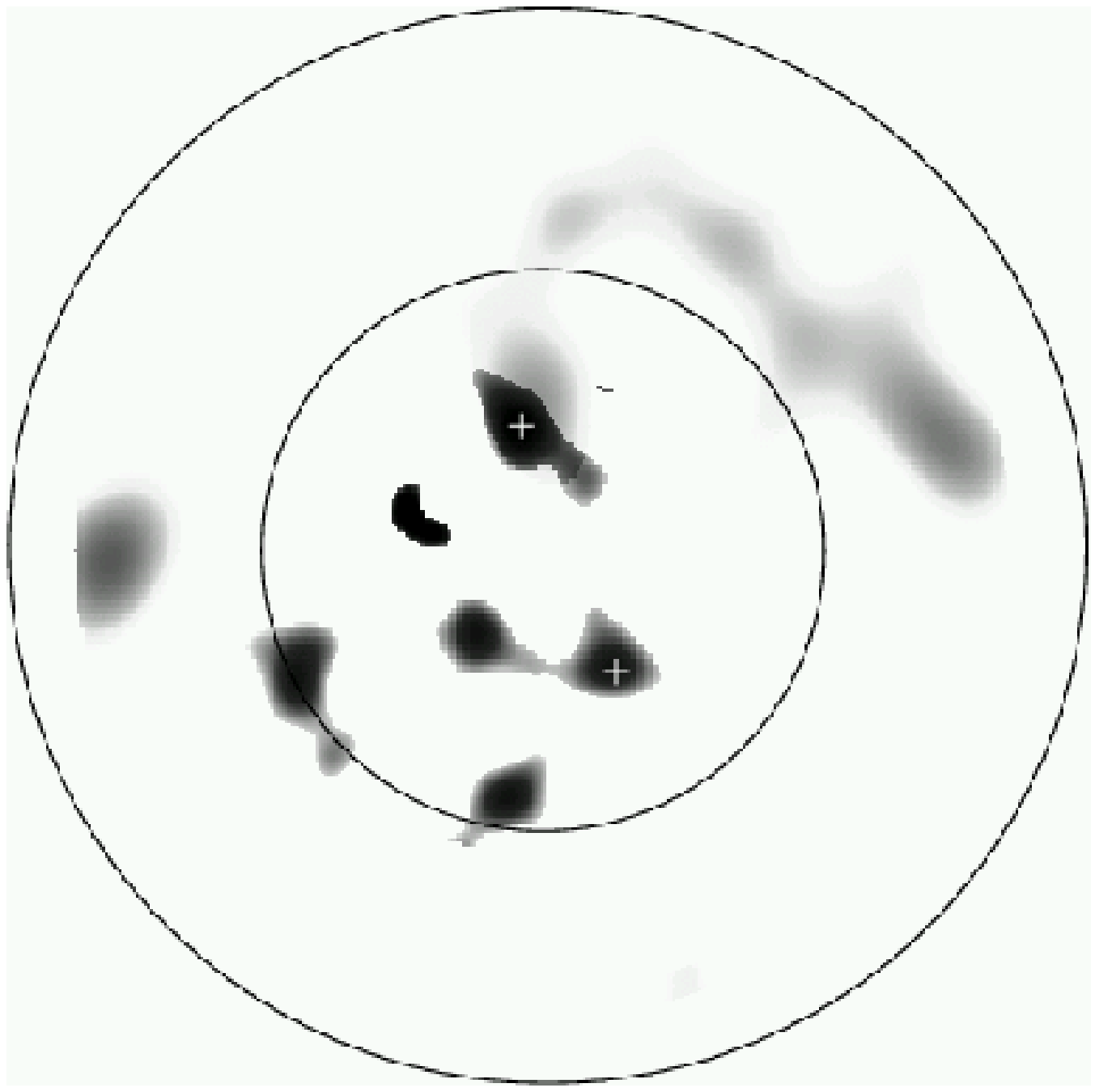}
\end{center}
\begin{center}
\includegraphics[angle=0, width=0.4\textwidth]{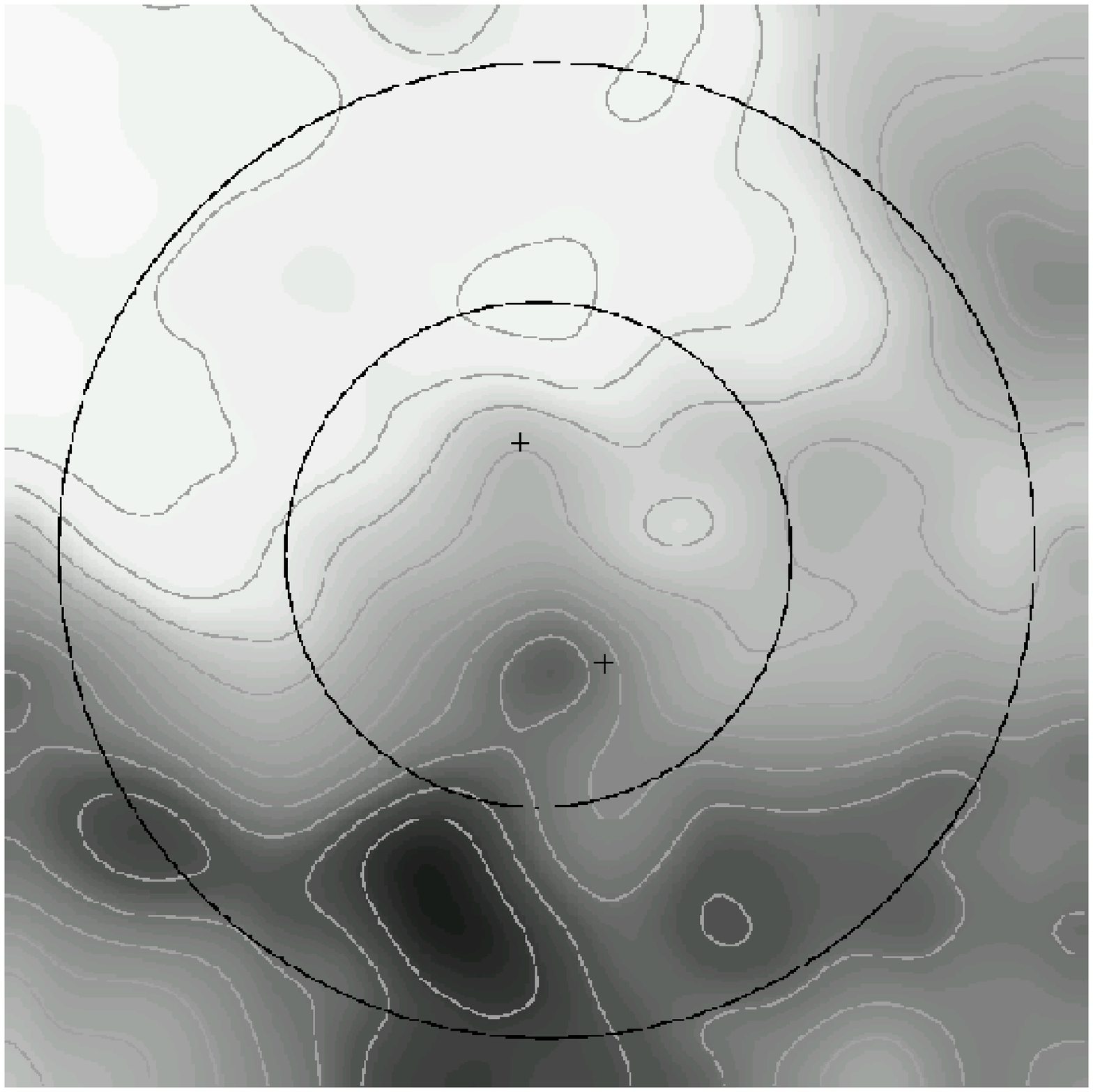}
\includegraphics[angle=0, width=0.4\textwidth]{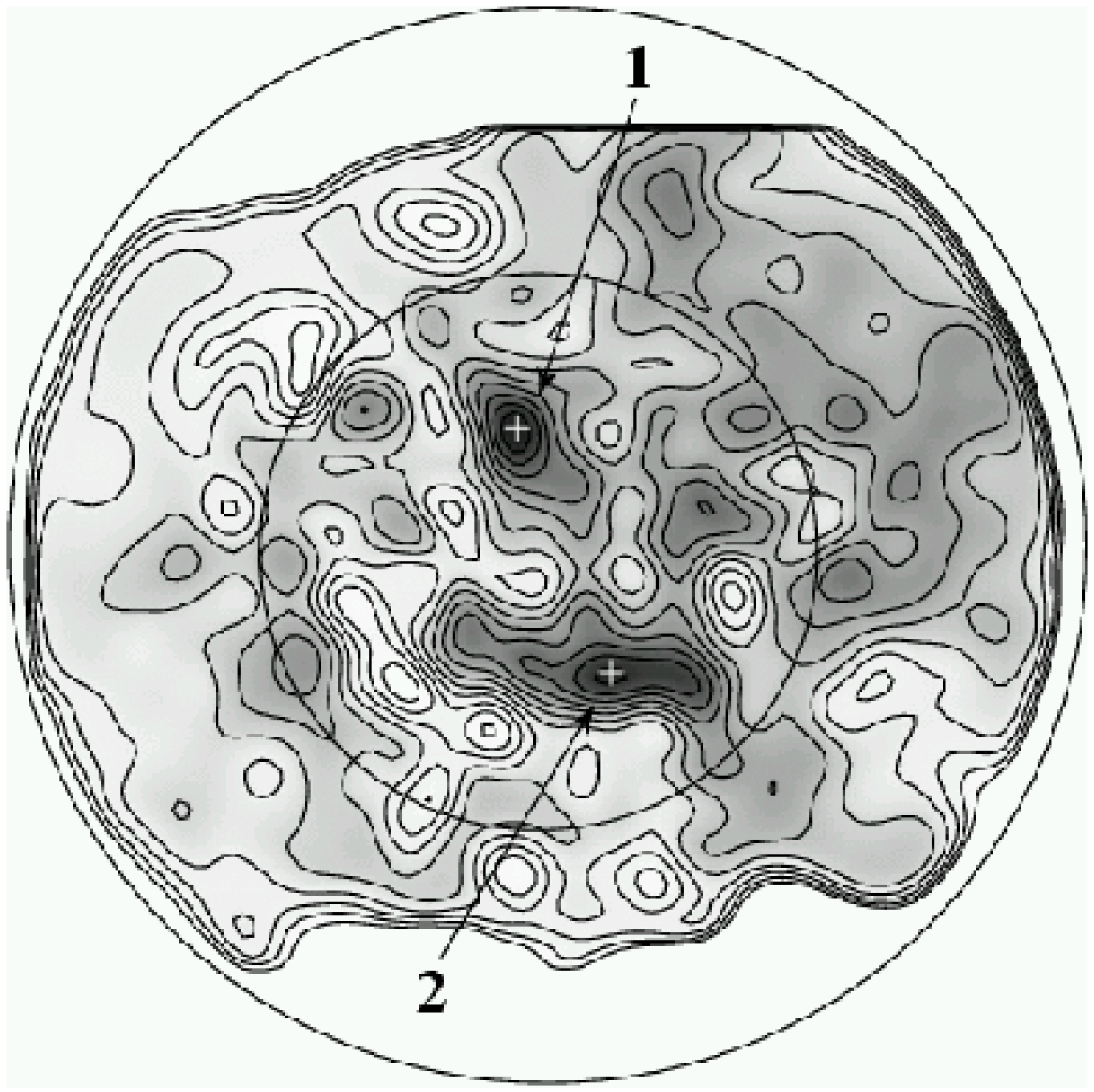}
\end{center}
\begin{center}
\caption{
 {\em (a)} Soft X-ray map smoothed to a resolution of 5.9
arcmin after removing the point sources identified in the
R2 and R6+7 maps; 
{\em (b)} Wavelet map showing extended
structures with high ($4\sigma$) statistical significance;
 {\em (c)} Map of Galactic HI column density (contours from
 5.3 to 10.7~10$^{19}$cm$^{-2}$, spaced by 0.6~10$^{19}$),
 lighter gray areas correspond to lower HI columns;
 {\em (d)} Soft map corrected for HI
absorption. As in Fig.\ref{maps_p}
the inner circle (1 deg in diameter) encloses
the region where the point source
removal is reliable. The two structures, marked with crosses, are in a
region where the ROSAT PSF is only $30\%$ worse than that in the 
center.}
\label{maps}
\end{center}
\begin{picture}(100,-100)
\put(70,525){{\em (a)}}
\put(280,525){{\em (b)}}
\put(70,310){{\em (c)}}
\put(280,310){{\em (d)}}
\end{picture}
\end{figure*}

\section{X-ray maps}\label{xmaps}

We have retrieved from the ROSAT archive the images of fields
contiguous to the Lockman Hole used by \citet{warwick} to create a large
scale soft X-ray map. These authors noted,
after subtraction of the point sources, large scale diffuse
structures in the $\sim 0.1-0.3 \rm{~keV}$ band. We have focused on one of the ROSAT
fields used by Warwick et al. and, more specifically, 
on "Field 4", centered at R.A.(J2000) $10^{\rm{h}}10^{\rm{m}}14^{\rm{s}}$
and DEC.(J2000) $+51^{\circ}45^{\prime}00^{\prime\prime}$, obtained
with a PSPC integration of 20 ksec.
We extracted the images in the various ROSAT bands\footnote{In the
following we will refer to such broad bands energy using the
conventional definitions reported in \citet{Snowden}} by means of
the software described in  \citet{Snowden}, which is optimized
for the analysis of extended structures in the ROSAT maps. 
In this way accurate maps of the background and of the
effective exposure over the whole field of view can be obtained. We mostly
focused on the soft image obtained in the R2 band (PI channels 21 to
40), which samples
the energy range $0.14-0.28 \rm{~keV}$. However, we also 
use the harder bands (R4+5, $0.44-1.21 \rm{~keV}$; R6+7, 
$0.73-2.04 \rm{~keV}$), both for the point sources removal 
and to constrain the nature of the diffuse emission. 
On the R2 image obtained with Snowden's code
the detection of extended structures has been performed with two different
methods:

\begin{enumerate}
\item We have constructed a map of the extended structures with an 
algorithm based on {\em SExtractor} \citep{bertin}. Our aim is to 
identify and subtract the point like sources which resolve most of the 
0.5-2.0 keV background.
This algorithm identifies the point sources
by increasing the size of the detection filter with the distance from
the center to match the broadening of the ROSAT PSF with the off--axis
angle. The detected
point sources in R2 and R6+7 bands are subtracted from the
soft image and replaced with the surrounding background.

Fig.~\ref{maps_p} shows the images, before the point sources removal and  
preliminarily smoothed with a gaussian kernel of 15 arcsec (i.e. one
PSPC pixel), 
in the soft (R2; Fig \ref{maps_p}{\em a}) and the hard (R6+7; 
Fig \ref{maps_p}{\em b}) bands.
The soft image was then smoothed with a gaussian kernel of 5.9
arcmin FWHM\footnote{This specific smoothing width was chosen 
because this is the minimum size of structures found with the
second method. A FWHM of 5.9 arcmin corresponds to gaussian with a
$\sigma$ of 10 pixels.} and is shown in Fig. \ref{maps}{\em a}.
The two main structures around the center of the field are
statistically significant to a level of $\sim 8 \sigma$ (assuming 
Poisson statistics).

\item To assess the significance of the structures found with the previous 
method we have also used a 
wavelet algorithm developed by \citet{Vikhlinin}. The aim of the
wavelet transform in this context is the complete background
subtraction to emphasize structures of certain angular size. The idea 
is to convolve an astronomical image with a kernel that is the 
difference of two gaussians, the first with sigma {\em a}, the second
subtracted from the previous with sigma $\mathit b=2a$. In the following
the kernel sizes will be reported as {\em a}.
We require the extended emission to be detected at a level 
of 4$\sigma$, with respect to the background fluctuations, and each detection
followed down to the 1$\sigma$ level. Given the strong radial
variation of the ROSAT PSF, not implemented into the wavelet algorithm,    
the structures detected with wavelet
kernel size less than $0.^\prime 75$ were considered as point
sources, while those detected by wavelet filters larger
than $1.^\prime 00$ are considered as extended. Features detected by wavelet filter
sizes between $0.^\prime 75$ and $1.^\prime 00$ are to be considered point
sources in the outer parts of the ROSAT field (and therefore
subtracted), whereas they were
considered to be potentially extended in the inner parts. 
In Fig. \ref{maps}{\em b} we show the map of 
structures detected with wavelet kernel
sizes in the range $0.^\prime 75-1.^\prime 25$ for the soft ROSAT band R2.
The advantage of the wavelet detection
algorithm is that it provides a quantitative measure
of the statistical reliability of detected extended structures with
respect to background fluctuations. The size measured for the smallest 
diffuse structure is $\sim 5.9 \rm{~arcmin}$.
\end{enumerate}
The most prominent diffuse/extended structures are in common to both
methods at least in the central part, where the detection of point sources is
more secure, indicating that they are not artifacts of the specific
algorithm. We mark as ``1'' and ``2'' in Fig. \ref{maps}{\em d} the two
most prominent structures in the central region.
In the outer parts, where the ROSAT PSF is much wider, the
identification of the extended structures is less reliable; however, 
even in these outer parts there is reasonable agreement between the maps
obtained with the two methods. 
Finally, as already commented on by Warwick et al 1998,
 we note that the same diffuse
structures are found by using other offset ROSAT pointings which
partly overlap this field,
lending
further weight to the argument that they are not instrumental artifacts.

\begin{figure}[!]
\begin{center}      
\vskip2truecm
\caption{Map of the extended hard emission (R6+7 band) obtained by
removing the point sources and smoothing to a resolution of 5.9 arcmin.}
\label{maphard}
\end{center}
\end{figure}
\begin{figure}[htbp]
\begin{center}      
\includegraphics[angle=0, width=0.5\textwidth]{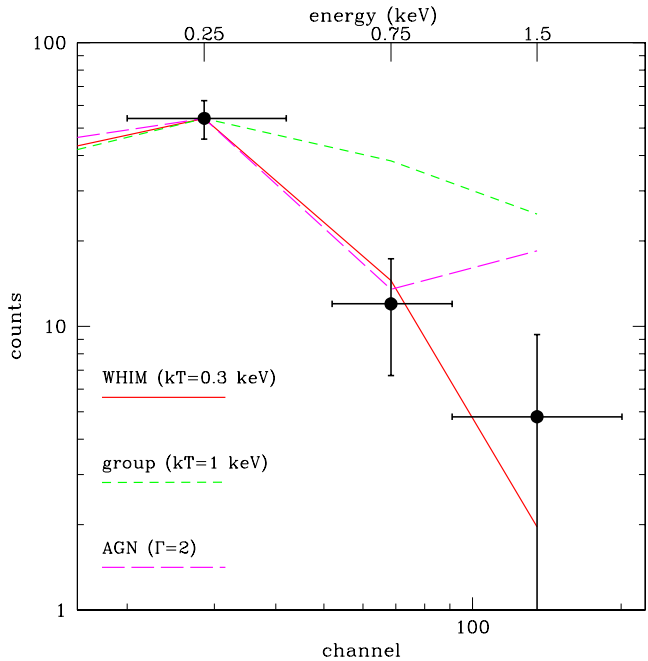}
\caption{The spectral shape of the strongest filament (marked ``1'' in
Fig.1{\em d}) compared with that expected from WHIM and groups of
galaxies at $z = 0.45$ and type 1 AGNs (see Sect.\ref{xmaps}).
The points are the measurements obtained by maps in R2, R4+5 and R6+7
band by extracting the total counts within an aperture of 7 arcmin around
the peak.}
\label{sxrb}
\end{center}
\end{figure}

In Fig.~\ref{maphard} we show the map of the
diffuse emission detected in the hard band (R6+7) obtained by removing
the point sources detected with {\it SExtractor} and by smoothing to a
resolution of 5.9 arcmin (this is the hard-band 
analogous of Fig.~\ref{maps}a).
With the exception of structure ``1'', which shows some faint
hard emission as discussed below,
the diffuse/extended
emission observed in the hard band (R6+7)
is not correlated with the diffuse
emission structures observed in the soft R2 band.
This indicates that the
diffuse features observed in the soft map are related to structures
with a much softer emission than that due to clusters or AGNs and, in
particular, cannot be ascribed to unresolved discrete sources which
make the 0.5-2 keV background.  \\A comparison between soft and medium (R4+5)
maps shows a weak correlation as expected from the hard tail of the
diffuse WHIM emission.  

In Fig.~\ref{sxrb} we plot the
fluxes\footnote{The uniform contribution from the Local Bubble
\citep{snowdenLHB} was subtracted from all images and the soft X-ray
emission was corrected for Galactic absorption as discussed in
Sect.\ref{HI}.}  in three bands (R2, R4+5, R6+7) of the brightest
diffuse structure (marked as ``1'' in Fig. \ref{maps}{\em d}), and the
expected spectra of various classes of sources: type 1 AGNs making the
0.5--2 keV background, groups of galaxies at $\rm{T}= 10^7 ~K$ (clusters
would be even hotter), and 
thermal emission by the WHIM at $\rm{T}=3 \times 10^6 ~K$. 
For both the groups and the WHIM we
adopted a 0.3 solar abundance and $z=0.45$ (see Sect. 
\ref{photom z} for the specific choice of this redshift). For
AGNs we assumed a power-law with $\Gamma = 2$. The choice of $\Gamma$ is lower
than the canonical $\Gamma = 2.5$ obtained from shallow ROSAT
surveys \citep{walter,brinkmann}, but is typical of deeper surveys 
\citep{mittaz} like ours. Furthermore it is similar to
the mean spectrum of the point sources in our maps 
($\Gamma = 2$ for point sources selected in R2 map
and $\Gamma = 1.7$ for those selected in R6+7 map). 
Fig.~\ref{sxrb} shows that the detected emission is 
inconsistent with emission by clusters or unresolved AGNs, but
is fully consistent with thermal emission from the WHIM.
It also indicates that the diffuse emission cannot 
be ascribed to possible residual wings of the PSF of point 
sources (AGNs) after their subtraction.

The one reported here is one of the first detections of WHIM 
at intermediate redshift. The diffuse
X-ray emission observed in our ROSAT maps can be compared with the
(projected) emission expected by the cosmological simulations of a WHIM
in the soft X-ray (e.g. Croft et al. 2001).
We used simulated images in the $0.2-0.3 \rm{~keV}$ band,
kindly provided to us by R.~Croft (private communication), and smoothed
them to the angular resolution of our
ROSAT maps (Sect. \ref{xmaps}). The average emission in the simulations is
$\langle F_{0.2-0.3keV}\rangle \approx 7.5 \times 10^{-13}
\rm{~erg~cm^{-2}s^{-1}deg^{-2}}$ with 1$\sigma$
fluctuations of about $2 \times 10^{-13} 
\rm{~erg~cm^{-2}s^{-1}deg^{-2}}$.
The average diffuse emission in the central region of our ROSAT map
is $10^{-12} \rm{~erg~cm^{-2}s^{-1}deg^{-2}}$ (extrapolated in the $0.2 - 0.3 
\rm{~keV}$ band), which is in good agreement with the simulations.
Taken in conjunction the results of the analysis described above strongly
imply that the diffuse structures described here represent one of the
first potential detections of a WHIM at intermediate redshift.

At a redshift of $\sim$0.45 (sect.5) the linear extension of
the brightest diffuse structure (1) corresponds to a physical size
of about 7~Mpc\footnote{Assuming $H_0=70$, $\Omega = 1$, $\Lambda =0$.}.

\section{Galactic HI maps}\label{HI}

The very soft R2 band ($0.14-0.28 \rm{~keV}$), where the diffuse
structures are more prominent, is strongly affected by even a low
amount of absorbing gas along the line of sight. As a consequence,
even though our field is close to the Lockman Hole \citep{lockman} 
where the Galactic 
absorption has a minimum ($\rm{N}_H \sim 5 \times 10^{19} 
\rm{cm^{-2}}$), a major concern is that the observed
structures might result from inhomogeneous distribution of Galactic HI
clouds along the line of sight. To exclude this possibility we
observed ``Field 4'' at 21cm with the Effelsberg 100~meter radio
telescope during October 2001.

This telescope has a HPBW (half power beam width) of $9^\prime$ at 21~cm,
which is close to the resolution of our ROSAT maps.
We used the 18-21~cm HEMT 2 channel receiver (T$_{\rm sys}\,\sim 25$~K) and
the 1024 channel autocorrelator\footnote{ The newer 8192 channel 
autocorrelator was not used due to interference problems at 21~cm.}.
The frequency resolution was set to $1.29 \rm{\;km\,s^{-1}}$ per channel
giving a
complete velocity coverage of $-380 \rm{\;km\,s^{-1}}$ to $+280 
\rm{\;km\,s^{-1}}$. This should
cover the HI emission from all Galactic and most HVC sources. 
The target fields were observed in a raster
pattern with spacing $4\farcs 3$ to Nyquist sample the beam.
Frequency switching was used with a 60~sec integration 
at each raster position (i.e. an effective integration time of 30~sec
per raster position).
The weather was exceptionally clear for most of the run.

The raw data were initially calibrated
using the CLASS and SPEC2 software,
converted to T$_{\rm sys}$, and then corrected for stray radiation
using the standard Effelsberg data reduction procedure
\citep{kalberla1,kalberla2}.
Observations of the Galactic reference field S7
(taken every 2 to 2.5 hrs) were then used to set
the flux scale.
Each raster spectrum was first Hanning smoothed to half its resolution and then
the `baseline', i.e. the shape of the continuum in each spectrum, was subtracted
by using a fourth order polynomial fit.
The HI emission between $-100 km\,s^{-1}$ and $66 km\,s^{-1}$ were
summed to produce the total HI emission in K km/s at each raster
point.  This value was multiplied by 1.8 $\times$ 10$^{18}$ in order
to obtain the total HI column in cm$^{-2}$ towards each raster point
(i.e. we assumed the HI emission is optically-thin).

In Fig.~\ref{maps}{\em c} we show the resulting N$_{H}$ map.
There is no correlation between the soft X-ray emission
and the distribution of N$_{HI}$ on the scales
of $\sim 10-20$~arcmin, indicating that the diffuse structures
observed in the former are not due to patchy HI absorption.
Some weak anticorrelation is present on the large scales between the smooth
North-South gradient of N$_H$ and the very extended X-ray emission
(correlation coefficient r~$\approx -0.3$); this anticorrelation is dominated
by the southern $\sim 30''$ of the field, i.e. outside the central
region which we are investigating, where a significant HI absorption
is present. If this southern part is removed the
anticorrelation disappears (r~$\approx + 0.14$).
The HI map was used to correct the R2 X-ray map for
absorption; the absorption-corrected map is shown in Fig.~\ref{maps}{\em d}.

\section{The spatial distribution of galaxies in the field}\label{galaxies}

To map the galaxy distribution we have imaged a subsection of the
ROSAT field in the optical. Indeed,
if the diffuse emission in the ROSAT maps is due to WHIM, then a
spatial correlation with galaxy overdensities is expected.

The optical data were obtained with the
Wide Field Camera (WFC) at the Isaac Newton Telescope in service
mode on 23 May 2000 and on 15 March 2001. This camera has
4 CCDs covering
a field of $\sim 34^\prime \times 34^\prime$ arcmin (with a gap of 
$\sim 10^\prime \times 10^\prime$
arcmin due to the distribution of the CCDs), in five broad band filters:
U (RGO), B (Kitt Peak), V (Harris), $r^\prime$ (Sloan) and 
$i^\prime$ (Sloan). Two fields
in the central region of the ROSAT map were observed; the
location of the two WFC pointings is shown in Fig.  \ref{pointings}:
\begin{figure}[htbp]
\begin{center}      
\vskip2truecm
\caption{Position of the WFC pointings over the R2 map.}
\label{pointings}
\end{center}
\end{figure}
the northern
field (field A hereafter) was observed on March-2001,
while
the southern field (field B hereafter) was observed on
May-2000. The seeing
was about 1.3$''$ and 1.6$''$ during observations of field A and field B, 
respectively.

\begin{figure*}[htbp]
\begin{center}
\includegraphics[angle=0, width=0.47\textwidth]{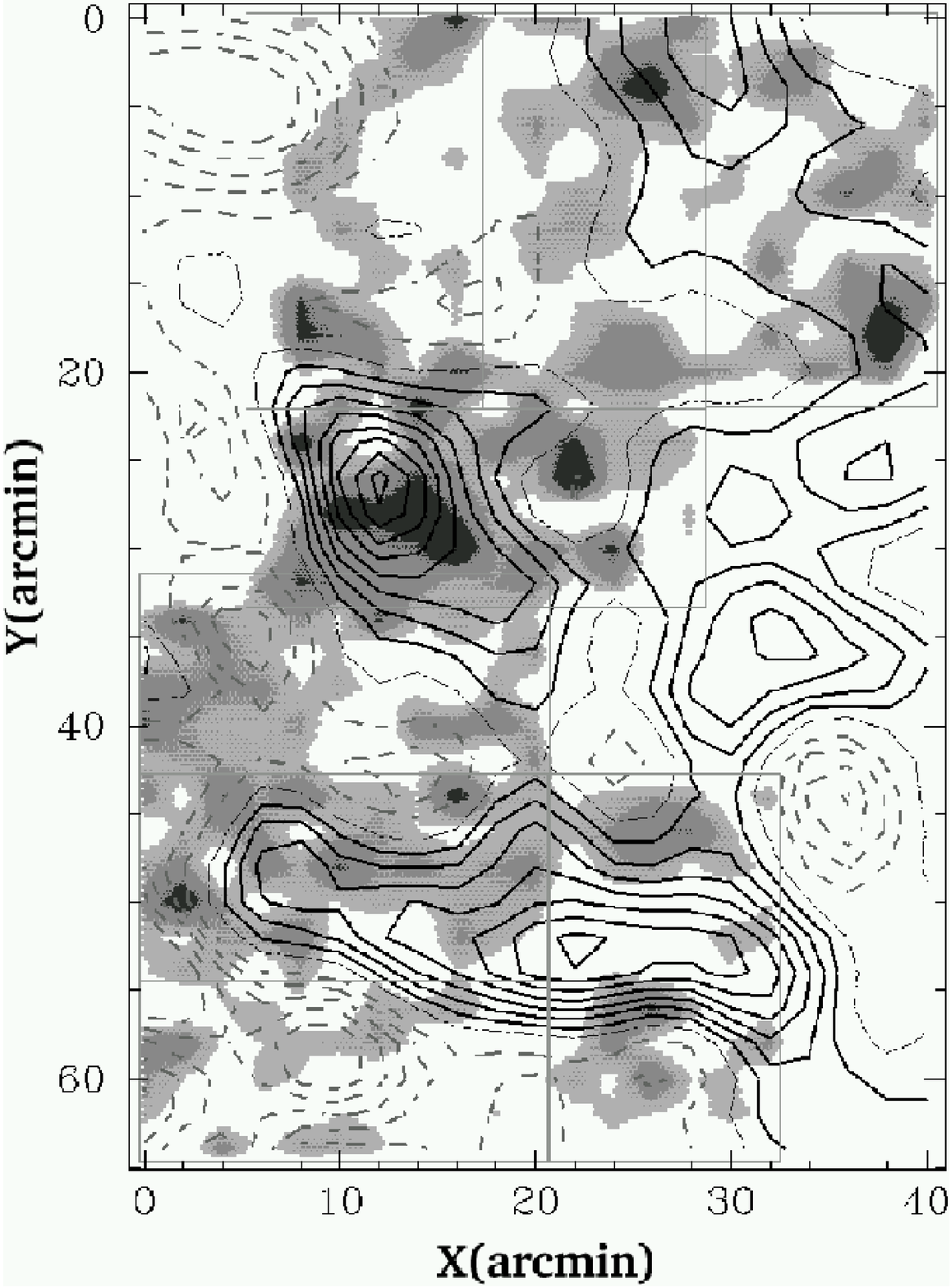}\rule[-2pt]{20pt}{0pt}\includegraphics[angle=0, width=0.47\textwidth]{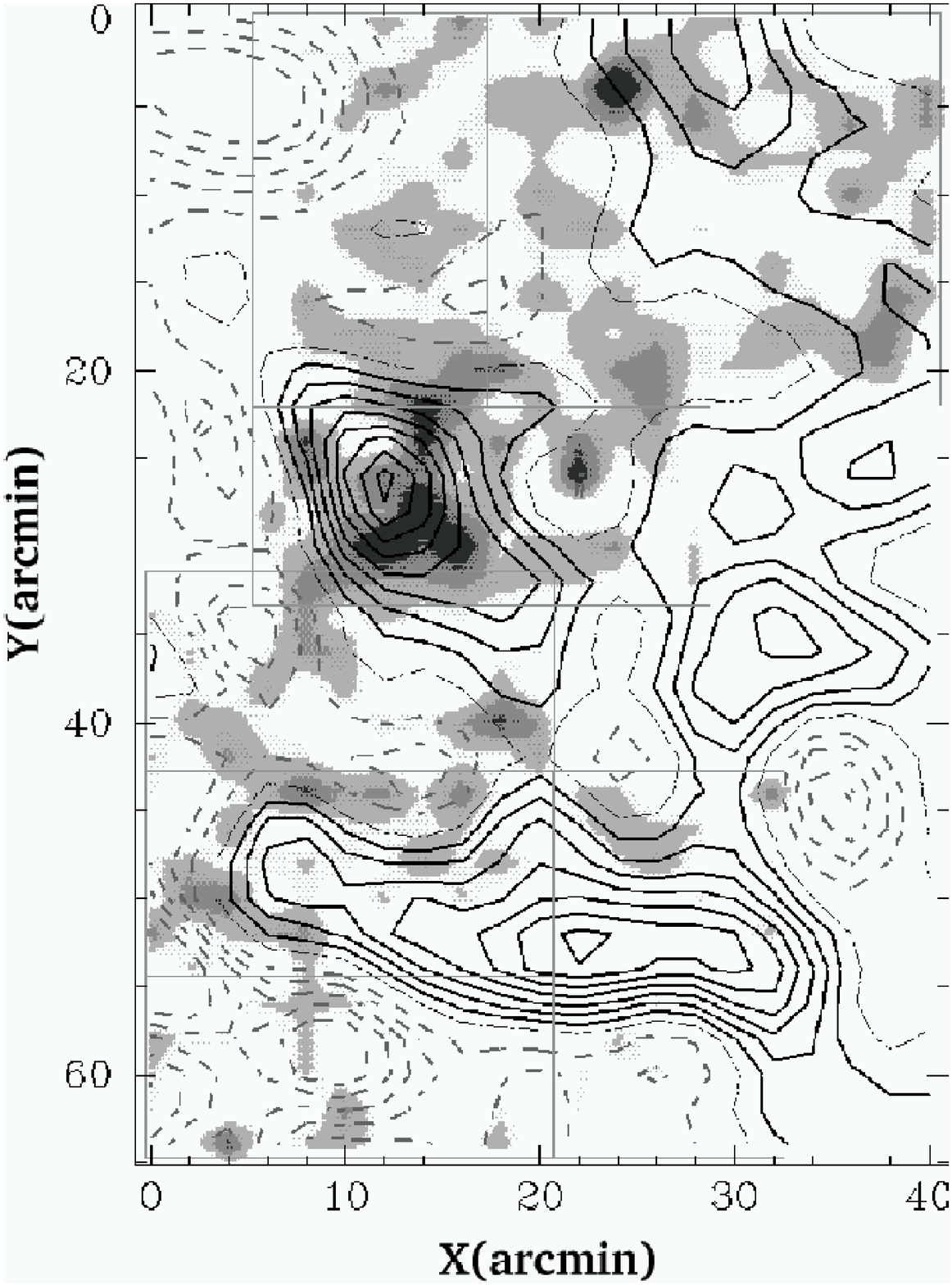}
\caption{The projected density of galaxies (grey scale) over the
central region of the R2 ROSAT field (panel {\em (a)}). The grey scale
delineates regions from just above the mean level of galaxy projected
density ($4.1 \rm{~galaxy~arcmin^{-2}}$; lightest grey) to $ > 7.5
\rm{~galaxy~arcmin^{-2}}$ (darkest grey). The diffuse soft X-ray
emission from Fig. \ref{maps}{\em (f)} is overlaid as contours. The
thick solid contours, thin dot--dashed contours and thick dashed
contours delineate regions with extended X-ray emission above, at, and
below the mean X-ray emission ($\rm 6~10^{-3}~counts~s^{-1}arcmin^{-2}$)
respectively. The
thin solid grey lines show the pattern of the wide field pointings. 
In panel {\em (b)} the projected galaxy density is restricted to
objects with measured photometric redshift ($\tilde{\chi}^2 <2$) and
$0.3 < z_{ph} < 0.6$. The mean galaxy projected density are at $1.4
\rm{~galaxy~arcmin^{-2}}$ (lightest grey) while the high density
regions (darkest grey) are above $2.8 \rm{~galaxy~arcmin^{-2}}$.}
\label{density_map}
\end{center}
\begin{picture}(0,0)
\put(5,415){{\em (a)}}
\put(265,415){{\em (b)}}
\end{picture}
\end{figure*}
For each field we have performed standard procedures of de-biasing,
flat fielding, correction of non linearity
and, in the case of $i^\prime$, correction for fringes.
We used SExtractor to build a preliminary
galaxy catalog. The
algorithm was run on the $r^\prime$
images as they are the most sensitive, at least for galaxies in the redshift
range of interest. We have detected 10877 objects in field A and
8650 in field B.\\
The zero points were determined by the observation of the standard
Landolt fields sa104 and sa101 \citep{landolt}.
However, to achieve the maximum accuracy in the photometric
calibration we refined the zero points with color--color diagrams. This
method consists of plotting the brightest non-saturated stars (in
our fields those with $17 < \rm{m(r^\prime)} < 20$ and stellarity
class\footnote{See \citet{bertin}} $> 0.95$)
in color--color diagrams to obtain the best agreement with the 
theoretical stellar main sequence. As reference we used the
digital stellar spectra compiled by \citet{pickles}.
The required corrections were below 5\%.\\
With the zero points refined in this way we re-ran SExtractor on the
images in all bands, for the objects detected in the $r^\prime$ band,
to derive the colors of all galaxies.

The $5\sigma$ limiting magnitude for point sources is about 23.6 in $r^\prime$.
Even though counts are 
consistent with those of SDSS \citep{sloan} and 
with \citet{metcalfe} up to $r^\prime=22.5$,
analysis of the differential
counts within the two fields shows that the completeness in field A is
different to that in field B. Field A
is up to 80\% complete at $r^\prime =23.3$ 
while field B reaches this completeness only at $r^\prime =22.9$. The 
number of objects are comparable only at magnitude 22.8. 
This difference most probably arises from different seeing conditions
prevailing during the observations.

We have quantified any possible galaxy overdensity associated with the
diffuse X-ray filaments by using two complementary methods:
1) mapping the projected density of objects and
2) searching for clumps of galaxies.

A map of the projected density of objects was constructed by grouping
all the objects in sub-sections of  $2^{\prime} \times 2^{\prime}$.
To avoid stellar contamination 
and to take into account the completeness of the galaxy catalog 
we have considered only those
objects in magnitude range $r^\prime = 19-22.8$.
The resulting galaxy density map is shown in grayscale
in Fig. \ref{density_map}{\em (a)}, 
where the contours
give the distribution of the diffuse soft X-ray emission
(from Fig. \ref{maps}{\em f}). 
Interestingly, the maximum galaxy density,  
about $10 \rm{\; galaxy/arcmin^2}$ (to be compared
with the average of $\sim 4.1 \rm{\; galaxy/arcmin^2}$), coincides
with the maximum of the soft X-ray emission. The probability of random
coincidence of the two maxima is $<1\%$ (inferred through simulation).
The second brightest filament (structure 2 in Fig. \ref{maps}{\em
f}) does not show an obvious associated overdensity of galaxies.

\begin{figure}[htbp]
\begin{center}
\includegraphics[angle=0, width=0.5\textwidth]{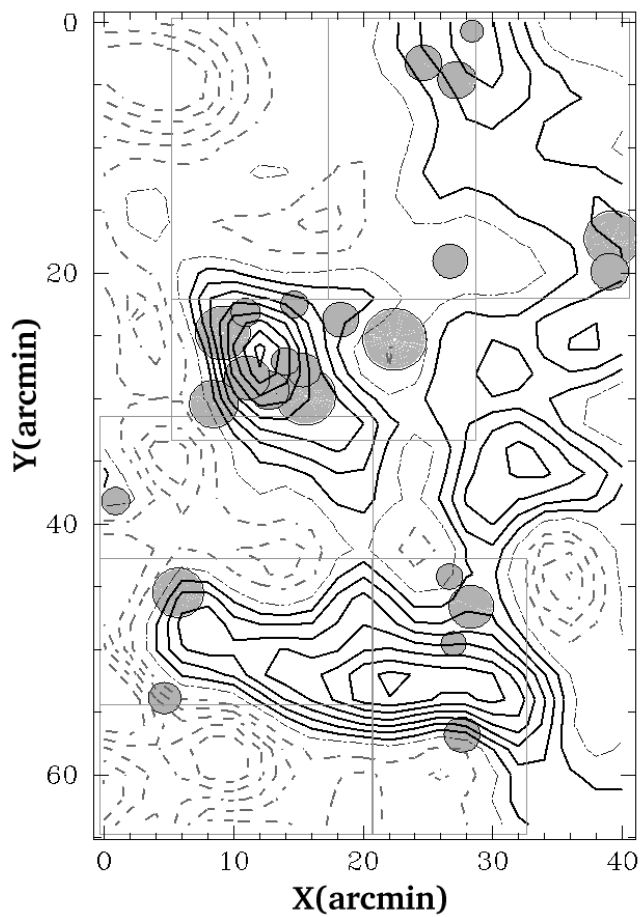}
\caption{Galaxy clumps found by Voronoi algorithm versus extended soft X-ray
emission (contours are the same of Fig. \ref{density_map}). 
Each association of galaxies is represented by a circle 
\citep[see][]{ramella}.  
Note the correlation between the galaxy associations and the structure
``1''.}
\label{clust_map}
\end{center}
\end{figure}

As an alternative method we have used the Voronoi tessellation
\citep{ramella} to detect overdensities of galaxies or clumps. 
By using a significance for the detection threshold of
75\% and choosing a confidence level for background fluctuations of
90\%, we found 24 clumps in the whole field (i.e. A$+$B).
The distribution and extension of these groups is shown in
Fig. \ref{clust_map}. 
There is a clear excess of clumps in the region of maximum X-ray
emission (structure 1) that confirms the result of the projected density map.
Some galaxy clumps are also detected around the southern
filament (structure 2) and around the NW (fainter) diffuse emission, but
the significance is obviously much lower.

Finally, we should mention that we have checked that the galaxies overdensities
obtained in our optical maps are not due to instrumental effects or to
statistical fluctuations. The former issue was tested by deriving
a map of projected density of stars, and establishing that they showed
no clustering or correlation
with the X-ray emission, nor with the Galactic HI maps, implying that
the association with the soft X-ray emission
is an intrinsic characteristic of the extragalactic objects.
Secondly, we run 1000
simulations of random distributions of galaxies with the
goal of identifying the degree of clustering due to statistical fluctuations.
Within the artificial data we found that "pixels" ($\rm 2'\times 2'$)
with galaxy densities in excess of 7~galaxy\,arcmin$^{-2}$
(the dark regions in Fig.~\ref{density_map}{\em (a)}) are less than 0.44\%
of the total distribution (upper limit at 1$\sigma$), while in the observation
the fraction of "pixels" with densities in excess of
7~galaxy\,arcmin$^{-2}$ is 8.4\% of the total.
Moreover, the "high density pixels" from the
simulations are distributed randomly and not clustered as observed in our
fields. The results of these simulations clearly indicate that the
overdensities observed by us are not due to statistical fluctuations.

\section{The 3-D distribution of galaxies in the field}\label{photom z}

In this section we focus on the overdensity of galaxies corresponding
to structure ``1''. 
We have derived galaxy photometric redshifts with the final
goal of determining whether the overdensity of galaxies observed in the
projected distribution is also observed in redshift space.
The photometric redshifts were obtained  by means of the {\em Hyperz}
code \citep{Bolzonella} and by exploiting the photometry
from all five available filters. Within the code we used 
all available templates
(namely ellipticals, spirals and irregulars), and we allowed for
a moderate degree of extinction, up to $\rm{A_v} = 0.6$. We checked that the
redshift distribution is not very sensitive to these
parameters. As previously stated, we limit our considerations to
galaxies with $r^\prime$ magnitudes between 19 and 22.8, which contribute up
to 54\% of
the entire catalog. Of these selected objects 59\% have redshift fitted with 
$\tilde{\chi}^2 < 1$, while
73\% are fitted with a $\tilde{\chi}^2 < 2$; in our analisys we have rejected 
objects with a photometric redshift fitted with a $\tilde{\chi}^2 > 2$.
In some cases the photometric redshift
implies an absolute magnitude brighter than $M_{r^\prime}=-23$; these are
mostly objects whose photometric redshifts are fitted with 
$\tilde{\chi}^2 > 2$ and the high luminosity is most likely a
consequence of an incorrect redshift determination. We have therefore 
discarded all objects with $M_{r^\prime}<-23$.
By imposing these constraints we are left with
63\% of the objects of the sample for the analysis of the photometric
redshifts.

\begin{figure}[htbp]
\begin{center}
\includegraphics[angle=0, width=0.45\textwidth]{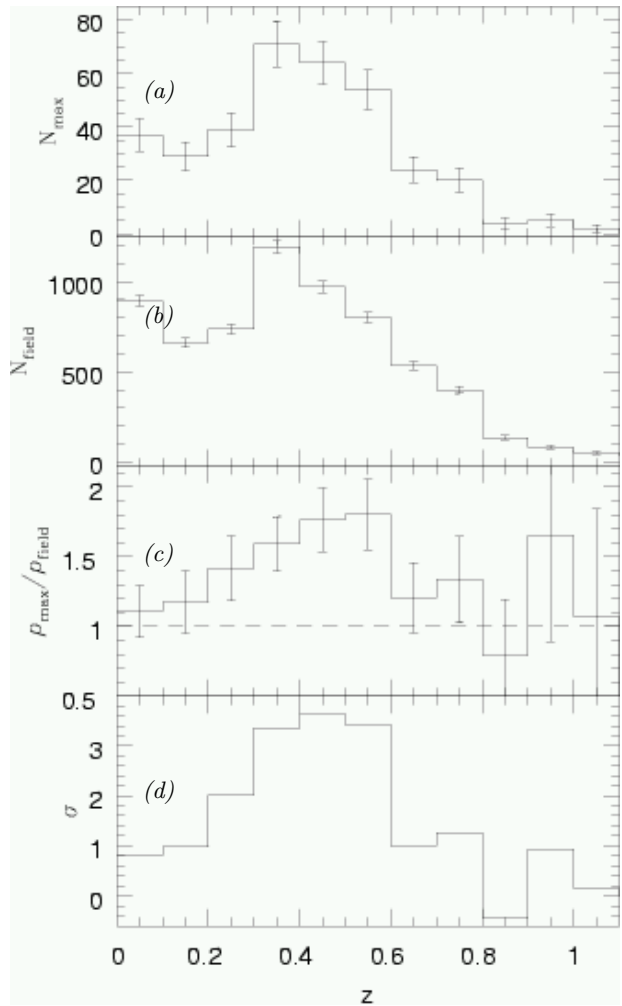}
\caption{Panel {\em (a)} Redshift distribution of the galaxies along the
$30\% \, \rm{FWHM}$ isophote in the region of maximum
X-ray emission; {\em (b)} The whole field distribution; {\em
(c)} Ratio
between the surface densities of the two regions. 
In case of no overdensity the histogram should
follow the dashed line, instead
a clear excess of objects in the redshift range 0.3$< z <$0.6.
{\em (d)} Excess of objects in units of $\sigma$ of the distribution.
}
\label{distrib_z}
\end{center}
\begin{picture}(0,0)
\put(60,465){{\em (a)}}
\put(60,380){{\em (b)}}
\put(60,290){{\em (c)}}
\put(60,200){{\em (d)}}
\end{picture}
\end{figure}

In the first two panels of Fig. \ref{distrib_z}
we show the distribution of the photometric redshifts for
galaxies within the isophote at 30\% from the X-ray maximum in structure "1"
(Fig.2), and the average redshift distribution
in the field. The ratio between these two distributions 
is shown in the third panel.
A clear excess of objects in the redshift range 0.3$< z <$0.6 is seen.
The excess has a high significance as shown in the fourth panel,
where the excess 
of galaxies in each redshift bin is reported in terms of $\sigma$ (the
latter inferred assuming Poisson statistics).
Given the expected errors in the photometric redshifts ($\sim \pm 0.15$)
the observed overdensity is consistent with a real structure at
$z \sim$0.45. By combining the three redshift bins where the overdensity
is observed, the significance of the redshift overdensity is $6 \, \sigma$.

In Fig. \ref{density_map}{\em (b)} we show the map of galaxy density, similar
to Fig. \ref{density_map}{\em (a)}, restricted to objects with
measured photometric redshift ($\tilde{\chi}^2 <2$) and in the redshift
range
0.3$< z_{ph} <$0.6. The contrast between density of galaxies on the
X-ray maximum relative to
the field is clearly higher than in the total projected
density map of Fig. \ref{density_map}{\em (a)}.

The association of the diffuse soft X-ray structure ``1'' with a 3-D
overdensity of galaxies, as inferred in the last two sections, strongly
imply that the former is associated to an extragalactic medium and,
more specifically, at a redshift of about 0.45.

\section{Summary and final remarks}\label{summary}

We have analyzed a deep ROSAT field in a region of high Galactic latitude
and low Galactic HI absorption. After removal of the point sources and
after correction for absorption, the softest map show evidence for
diffuse/extended structures. These diffuse structures have an X-ray
energy distribution that is much softer than expected from clusters, groups or
unresolved emission from AGNs, but it is consistent with emission from
the diffuse warm intergalactic medium
expected by the cosmological models, both in terms of shape
(plasma at kT$\sim$0.3~keV) and of flux ($\rm \langle F_{0.2-0.3keV}\rangle
\approx 10^{-12}~erg~cm^{-2}s^{-1}deg^{-2}$).

To discriminate between a Galactic or extragalactic nature of the
diffuse gas we have correlated the soft X--map with multiband optical
images in this field.
The most prominent diffuse X-ray structure in the ROSAT map appears
associated with an overdensity of galaxies at a photometric redshift of
$\sim 0.45$. This association, along with the X-ray
properties of the former (sect.\ref{xmaps}), strongly suggest that
we are observing an extragalactic structure most likely tracing
the warm intergalactic medium predicted by cosmological theories,
in this case at redshift $\sim 0.45$, which is expected to be the main
reservoir of baryonic matter at low redshifts.

The second most prominent diffuse
X-ray structure ``2'' is not associated, in our optical maps, with a pronounced
galaxy overdensity, although there are some clumps of galaxies
surrounding it (Fig. \ref{clust_map}).
Either this structure is not extragalactic (e.g.
associated to our Galactic halo) or it is associated with galaxies at
redshift higher than 0.8, where most galaxies escape detection in our
optical images. Finally, this may be a case of warm baryonic
gas at relatively low redshift, not enclosing galaxy overdensities,
but possibly bridging those few galaxy clumps detected in its surrounding.

Multi-object spectroscopy should provide
a critical test on the nature of these diffuse structures.
Indeed the specroscopic information would both
confirm the galaxy redshift distribution and give an estimate of the
involved virial masses to be compared with the temperature
inferred for the barionic gas.

\begin{acknowledgements}

We are grateful to R.~Croft for providing us with the unpublished
maps of his simulations and to M.~Bolzonella for making available
her Hyperz code and for providing additional information and suggestions.
We thank G. Hasinger for helpful
comments during the early stages of this work.
We are also grateful to the staff of the Isaac Newton Telescope and of the
Effelsberg 100 meter telescope for their support during the observations.
We are grateful to Peter Kalberla for giving us access to
Effelsberg calibration software and Alison Peck for help
with the observations.

\end{acknowledgements}


\end{document}